\documentclass[rnaas]{aastex62}

\pdfoutput=1

\usepackage{lmodern}
\usepackage{microtype}
\usepackage{url}
\usepackage{amsmath}
\usepackage{amssymb}
\usepackage{natbib}
\usepackage{multirow}
\usepackage{graphicx}
\renewcommand{\eqref}[1]{\ref{eq:#1}}
\newcommand{\Eq}[1]{Equation~(\eqref{#1})}

\newcommand{\eqlabel}[1]{\label{eq:#1}}

\begin{document}\raggedbottom\sloppy\sloppypar\frenchspacing

\title{%
A robust method to measure centroids of spectral lines
}

\author[0000-0003-1534-5186]{Richard Teague}
\affil{University of Michigan, Ann Arbor, MI}

\author[0000-0002-9328-5652]{Daniel Foreman-Mackey}
\affil{Flatiron Institute, New York, NY}


\section{Background}

When analyzing data cubes produced by radio telescopes or integral field units,
astronomers frequently aim to produce a map of velocity as a function of pixel
coordinate or time.
In some cases, the spectral dimension is dominated by a single line and the
operation of inferring a velocity is simplified to a question of determining
the centroid of the line.
For this problem, there exist multiple popular methods \citep[see, for
example][]{deBlok:2008}, however the most common are the ``intensity weighted
average velocity'' (also known as the ``first moment'' method) and the
velocity of the maximum intensity, called the ``ninth moment''\footnote{We
emphasize that this is a misnomer as it does not relate to any statistical
moment.}, as they are implemented by \texttt{CASA} \citep{McMullin:2007}.
Even when the emission is dominated by a single line component, these methods
do not offer the accuracy or precision required to extend studies to small
scale structures.

The ``first moment'' approach is strongly affected by the noise in the spectrum
and any asymmetry in the line profile.
Sigma clipping is typically used to circumvent these issues, however this can
result in the addition of spurious features or the masking of astrophysically
interesting features in the resulting velocity map.
``Ninth moment'' maps are more robust to these issues, however the precision of
this approach is limited by the velocity resolution of the data.

Other approaches involve fitting a model line profile (e.g.~a Gaussian) to the
data, but these methods are both more computationally expensive and more
sensitive to the details of the line shape.
In the case of a double-Gaussian profile, due to radial infall, for example, the
line center for a single Gaussian component (or a higher order expansion) will
therefore be offset to account for this double component.

In this note we present a new method for inferring line centroids that is
simple, computationally efficient, and robust to noise and errors in line
shape.

\section{Technical details}

It has been demonstrated that the centroid of photometric sources can be
determined with near-optimal precision by fitting a quadratic surface to the
brightest central pixels of the point spread function \citep{Vakili:2016}.
Here we apply this method in one dimension to estimate the centroid of
spectral lines.
The fundamental idea is that we can fit a quadratic model to the brightest
pixel and one pixel on either side to estimate the curvature of the line
profile near maximum.

The procedure is as follows:
\begin{enumerate}

{\item We find the pixel of maximum intensity in the spectrum and extract that
pixel value $I_0$ and the one to the left $I_-$ and right $I_+$.}

{\item We model these three intensities as a quadratic function
\begin{eqnarray}
I(x) &=& a_0 + a_1\,(x-x_0) + a_2\,{(x-x_0)}^2
\eqlabel{quad}
\end{eqnarray}
where $x_0$ is the pixel coordinate of maximum intensity and $x$ is in units
of pixels.
Given the intensities in the three target pixels, we find
\begin{eqnarray}
a_0 &=& I_0 \\
a_1 &=& \frac{1}{2}(I_+ - I_-) \\
a_2 &=& \frac{1}{2}(I_+ + I_- - 2\,I_0) \quad.
\end{eqnarray}
}

{\item The maximum of \Eq{quad} is at
\begin{eqnarray}
x = x_\mathrm{max} &=& x_0 - \frac{a_1}{2\,a_2} \\
    &=& x_0 - \frac{f_+ - f_-}{2\,(f_+ + f_- - 2\,f_0)} \quad,
\end{eqnarray}
and this is our estimate for the centroid of the line in pixel coordinates.
}

\end{enumerate}

We can also estimate the statistical uncertainty on $x_\mathrm{max}$ by
linearizing and propagating the uncertainty from the fluxes to the centroid
estimate.
If we assume that the intensity uncertainties are normal, independent, and
homoskedastic with standard deviation $\sigma$, we find the
following approximation for the statistical uncertainty on $x_\mathrm{max}$
\begin{eqnarray}
{\sigma_{x_\mathrm{max}}}^2 = \frac{\sigma^2}{8}\,\left(
    \frac{3}{{a_2}^2} + \frac{{a_1}^2}{{a_2}^4}
\right)\quad.
\end{eqnarray}
We emphasize that this only includes the statistical uncertainty and any
application of this method to real data will also require a treatment of
systematic uncertainties\footnote{The code released
alongside this note includes an implementation and proof of more general noise
models at \url{https://github.com/richteague/bettermoments}.}.

\section{Demonstration of Method}

\begin{figure}[htbp]
\centering
\includegraphics[width=\textwidth]{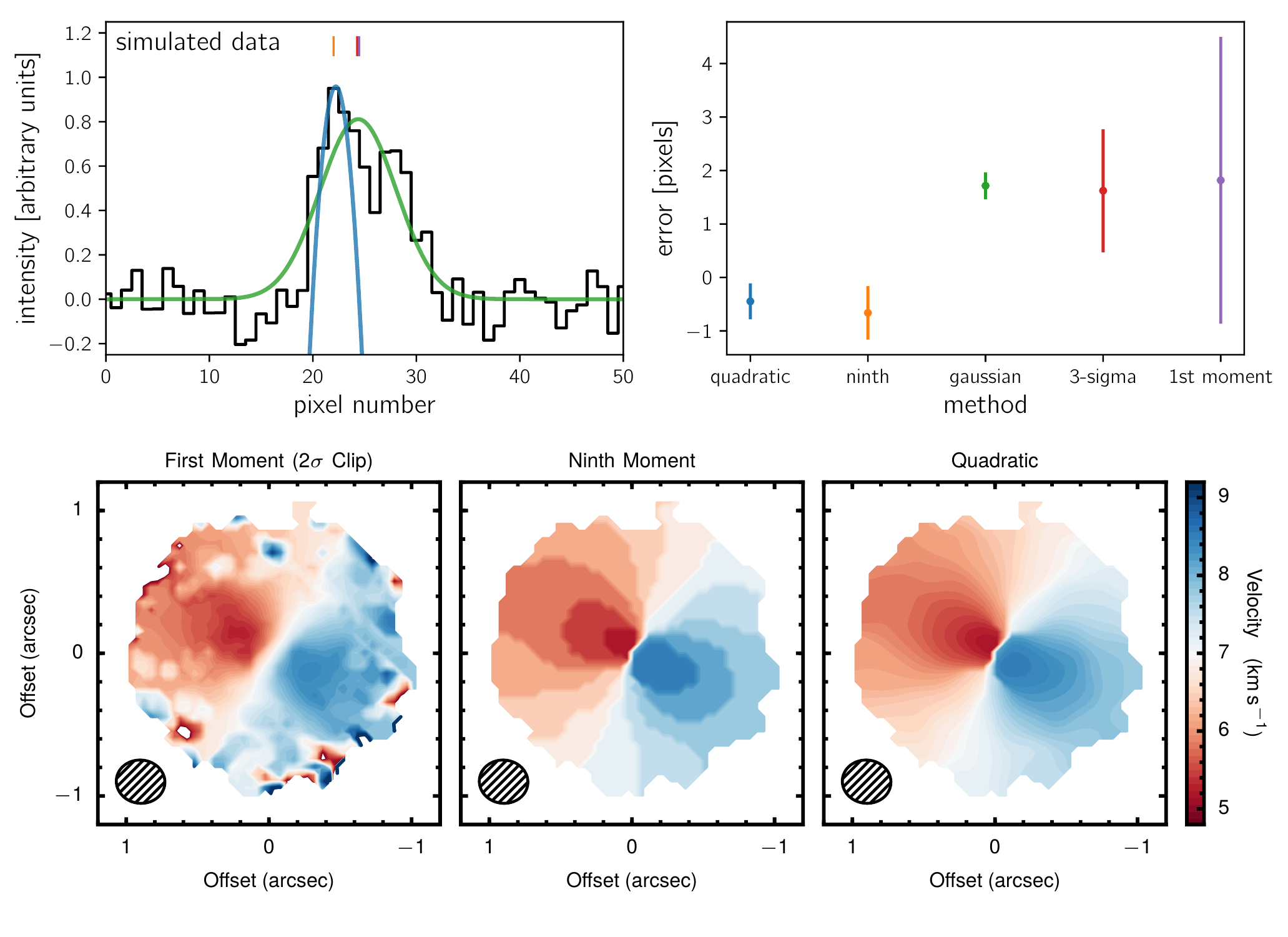}
\caption{%
Top row: Comparisons of different line centroid methods applied to simulated
data.
Left: A Gaussian fit to the data is shown in green and the quadratic fit to
the peak pixels is in blue.
Vertical ticks show the velocity inferred using the other approaches with the
same colors as in the right panel.
Right: The difference between the inferred centroid and the truth for the
methods listed along the x-axis.
The error bars indicate the statistical uncertainty of each method.
Bottom row: Application of the three methods to observations of HD~135344B.
\label{figure}}
\end{figure}

The top row of Figure~\ref{figure} compares different methods for estimating
a line centroid the case of an asymmetric line profile, modeled as the sum of
two Gaussian components.
Both the first moment and Gaussian fit results are biased by the asymmetry of
the line.
Although the ``ninth moment'' approach avoids this bias, its precision is
limited by the velocity resolution of the data.

The bottom row shows velocity maps of $^{13}$CO from the protoplanetary disk,
HD~135344B \citep[ALMA Project 2012.1.00158.S]{vanderMarel:2016}, imaged at a
velocity resolution of $330~{\rm m\,s^{-1}}$.
With a $2\sigma$ clip applied to the data, a ``first moment'' approach results
in a noisy map, making it hard to distinguish real structures from noise.
As with the asymmetric spectrum, the ``ninth moment'' is less affected by the
noise but the resulting map loses fidelity due to the coarse velocity
resolution of the data.
The quadratic approach does not suffer from noise features while maintaining a
high precision on the inferred velocity and achieving a statistical uncertainty
of $47~{\rm m\,s^{-1}}$, or 15\% of the velocity resolution.

The code used to generate this Figure and an open source implementation of
this method is available at
\url{https://github.com/richteague/bettermoments}.

\bibliography{bettermoments}

\end{document}